\documentclass[fleqn,12pt,twoside]{article}
\usepackage{espcrc1}

\usepackage{graphicx}
\usepackage[figuresright]{rotating}

\newcommand{\AmS}{{\protect\the\textfont2
  A\kern-.1667em\lower.5ex\hbox{M}\kern-.125emS}}

\title{Strangeness in the proton and $N^*(1535)$}

\author{B.S.Zou\address[MCSD]{Institute of High Energy Physics, CAS, P.O.Box 918(4),
Beijing 100049, China}%
        }

\begin{document}

\maketitle

\begin{abstract}
The newest progress on the study of the strangeness in the proton
and in the lowest negative parity nucleon excited state
$N^*(1535)$ is reviewed. Implications on the internal quark
structure of the proton, $N^*(1535)$ and other baryons are
discussed. The diquark cluster picture for the 5-quark components
in baryons gives a natural explanation not only to the empirical
indications for a positive strangeness magnetic moment $\mu_s$ and
positive strangeness radius of the proton but also the
longstanding mass-reverse problem of $N^*(1535)$, $N^*(1440)$ and
$\Lambda^*(1405)$ resonances as well as the unusual decay pattern
of the $N^*(1535)$ resonance. Evidence for possible existence of
$N^*(1535)$'s ${1/2}^-$ SU(3) nonet partners in this picture is
pointed out, and suggestion is made to search for these $1/2^-$
hyperon excited states under the well known $\Sigma^*(1385)$,
$\Lambda^*(1520)$ and $\Xi^*(1530)$ peaks in various reactions.
\end{abstract}

\section{Introduction}

In classical quark models, each baryon is composed of three
quarks. The simple 3q constituent quark model has been very
successful in explaining the static properties, such as mass and
magnetic moment, of the spatial ground states of the flavor SU(3)
octet and decuplet baryons. Its predicted $\Omega$ baryon with
mass around 1670 MeV was discovered by later experiments.

However its predictions for the spatial excited baryons failed
badly. In the simple 3q constituent quark model, the lowest
spatial excited baryon is expected to be a ($uud$) $N^*$ state
with one quark in orbital angular momentum $L=1$ state, and hence
should have negative parity. Experimentally \cite{PDG}, the lowest
negative parity $N^*$ resonance is found to be $N^*(1535)$, which
is heavier than two other spatial excited baryons :
$\Lambda^*(1405)$ and $N^*(1440)$. In the classical 3q constituent
quark model, the $\Lambda^*(1405)$ with spin-parity $1/2^-$ is
supposed to be a ($uds$) baryon with one quark in orbital angular
momentum $L=1$ state and about 130 MeV heavier than its $N^*$
partner $N^*(1535)$; the $N^*(1440)$ with spin-parity $1/2^+$ is
supposed to be a ($uud$) state with one quark in radial $n=1$
excited state and should be heavier than the $L=1$ excited ($uud$)
state $N^*(1535)$, noting the fact that for a simple harmonic
oscillator potential the state energy is $(2n+L+3/2)\hbar\omega$.
So for these three lowest spatial excited baryons, the classical
quark model picture is already failed.

The second outstanding problem in the classical quark model is
that it predicts a substantial number of `missing $N^*$ states'
around 2 GeV/$c^2$, which have not so far been observed
\cite{Capstick1}. The third outstanding problem is that from deep
inelastic scattering and Drell-Yan experiments the number of $\bar
d$ is found to be more than the number of $\bar u$ by 0.12 in the
proton \cite{Garvey}.

\begin{figure}[ht]
\vspace{0cm}
\hspace{0cm}\includegraphics[scale=0.47]{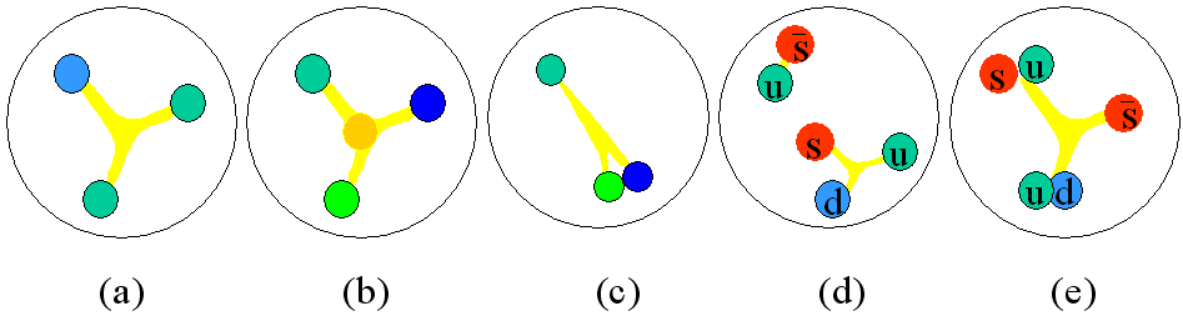}
\vspace{-0.5cm} \caption{Various pictures for internal quark-gluon
structure of baryons: (a) $3q$, (b) $3qg$ hybrid, (c) diquark, (d)
meson-baryon state, (e) pentaquark with diquark clusters. }
\label{fig0}
\end{figure}

The failure of the classical quark models raises a fundamental
question : what are effective degrees of freedom for describing
the internal structure of baryons? Several pictures based on
various effective degrees of freedom have then been proposed, such
as quark-gluon hybrid model, diquark model, meson-baryon state,
pentaquark with diquark clusters as shown in Fig.\ref{fig0}.

Among various pictures for the baryon, the meson cloud picture
seems quite successful. With this picture, the excess of $\bar d$
over $\bar u$ in the proton is explained by a mixture of $n\pi^+$
with the $\pi^+$ composed of $u\bar d$ \cite{Thomas}; the
$N^*(1535)$ and $\Lambda^*(1405)$ are ascribed as quasi-bound
states of $K\Sigma$ and $\bar KN$, respectively \cite{weise}.

To understand 5-quark components in baryons, to study the
strangeness in the proton and in the lowest negative parity
nucleon excited state $N^*(1535)$ should be very instructive. In
the following, by studying the strangeness in the proton and
$N^*(1535)$, we will show that instead of the conventional ``meson
cloud'' configurations the diquark-diquark-antiquark
configurations could play very important or even dominant role in
excited baryons.

\section{Strangeness in the proton}

There has long been some evidence that there may be $\bar ss$
pairs in the nucleon \cite{Ellis}. Several measurements including
the $\pi N$ $\sigma$-term, neutrino-induced charm production and
polarization effects in electron-nucleon deep-inelastic scattering
indicate that there may be significant $s\bar s$ component in the
proton \cite{Ellis,Alberico,Beck}. The excesses of $\phi$
production in $\bar pp$ annihilation \cite{Amsler} above the naive
OZI rule predictions were also used to argue in favor of a
significant $\bar ss$ component in the proton \cite{Karliner}
although the results can also be explained by two-step
contribution \cite{Locher} without introducing explicitly the
$\bar ss$ component in the proton.

For the strangeness in the proton, an interesting issue is whether
the $s$ and $\bar s$ distributions are the same ? In the
meson-cloud picture with a mixture of $K^+\Lambda$ component in
the proton, $s$-$\bar s$ asymmetry is naturally expected.  The
strangeness spin $\Delta_s$, strangeness magnetic moment $\mu_s$
and strangeness radius $r_s$ are all predicted to be negative
\cite{brodsky,musolf}. There are some empirical indications for a
negative $\Delta_s$ value as ($-0.10\pm 0.06$) \cite{Brad,spin},
which is compatible with the expectation from the simple
meson-cloud model.

For the strangeness magnetic moment $\mu_s$ and strangeness
radius, there are many other model predictions, such as including
$K^*\Lambda$ meson-cloud contribution which may change the sign of
the $\mu_s$ by adjusting model parameters \cite{isgur}.

However, recently four experiments on parity violation in
electron-proton scattering suggest that both strangeness magnetic
moment $\mu_s$ and strangeness radius $r_s$ of the proton  are
positive \cite{mus-exp}. This is in contradiction with most
theoretical calculations \cite{zr,bijker}.

A complete analysis \cite{zr} of the relation between these
strangeness observables and the possible configurations of the
$uuds\bar s$ component of the proton concludes that, for a
negative $\Delta_s$, positive $\mu_s$ and $r_s$, the $\bar s$ is
in the ground state and the $uuds$ system in the $P$-state. The
conventional $K^+\Lambda$ configuration as shown in
Fig.~\ref{fig0}(d) has the $\bar s$ mainly in $P$-state and hence
leads to negative value for both $\mu_s$ and  $r_s$. The hidden
strangeness analogues of recently proposed diquark cluster models
\cite{jaffe} for the $\theta^+$ pentaquark as shown in
Fig.~\ref{fig0}(e) have $\bar s$ in the ground state and the
$uuds$ system in the $P$-state, hence give positive value for both
$\mu_s$ and  $r_s$. The diquark cluster configurations also give a
natural explanation for the excess of $\bar d$ over $\bar u$ in
the proton with a mixture of $[ud][ud]\bar d$ component in the
proton.

Some recent theoretical attempts with closer relation to QCD
\cite{Leinweber,jixd} have not given a conclusive view on the sign
of the $\mu_s$. A very recent analysis \cite{young} of combined
set of parity-violating electron scattering data gives the strange
form factors to be consistent with zero. If the result be further
proved by more precise measurements and analyses in the furure, it
could mean that there may be about equal amount of meson-cloud
components and $q^2q^2\bar q$ components in the proton.

\section{Strangeness in $N^*(1535)$ and implication on its $1/2^-$ SU(3) nonet partners}

From the study of the proton, we know that there should be at
least about $20\%$ mixture of the penta-quark components in the
proton to reproduce its large $\bar u$-$\bar d$ asymmetry ($\bar
d-\bar u\approx 0.12$) and $s$-$\bar s$ asymmetry. Then in the
excited nucleon states, $N^*$ resonances, more multi-quark
components should be expected. To understand the properties of the
$N^*$ resonances, it is absolutely necessary to consider these
multi-quark components.

Recently BES experiment at Beijing Electron-Positron Collider
(BEPC) has been producing very useful information on $N^*$
resonances \cite{ppeta,Yanghx,pnpi}. From BES results on
$J/\psi\to\bar pp\eta$ \cite{ppeta} and $\psi\to\bar pK^+\Lambda$
\cite{Yanghx}, the ratio between effective coupling constants of
$N^*(1535)$ to $K\Lambda$ and $p\eta$ is deduced to be
$g_{N^*(1535)K\Lambda}/g_{N^*(1535)p\eta} =1.3\pm 0.3$ \cite{lbc}.
With previous known value of $g_{N^*(1535)p\eta}$, the obtained
new value of $g_{N^*(1535)K\Lambda}$ is shown to reproduce recent
$pp\to pK^+\Lambda$ near-threshold cross section data as well.
Taking into account this large $N^*K\Lambda$ coupling in the
coupled channel Breit-Wigner formula for the $N^*(1535)$, its
Breit-Wigner mass is found to be around 1400 MeV, much smaller
than previous value of about 1535 MeV obtained without including
its coupling to $K\Lambda$.

The nearly degenerate mass for the $N^*(1535)$ and the $N^*(1440)$
resonances can be easily understood by considering 5-quark
components in them \cite{lbc,zhusl,riska2}. The $N^*(1535) 1/2^-$
could be the lowest $L=1$ orbital excited $|uud>$ state with a
large admixture of $|[ud][us]\bar s>$ pentaquark component having
$[ud]$, $[us]$ and $\bar s$ in the ground state. Note that the
$N^*$ with negative parity cannot have $|[ud][ud]\bar d>$
component with two identical diquarks. The $N^*(1440)$ could be
the lowest radial excited $|uud>$ state with a large admixture of
$|[ud][ud]\bar d>$ pentaquark component having two $[ud]$ diquarks
in the relative P-wave. While the lowest $L=1$ orbital excited
$|uud>$ state should have a mass lower than the lowest radial
excited $|uud>$ state, the $|[ud][us]\bar s>$ pentaquark component
has a higher mass than $|[ud][ud]\bar d>$ pentaquark component.
The large mixture of the $|[ud][us]\bar s>$ pentaquark component
in the $N^*(1535)$ may also explain naturally its large couplings
to the $N\eta$ and $N\Lambda$ meanwhile small couplings to the
$N\pi$ and $K\Sigma$. In the decay of the $|[ud][us]\bar s>$
pentaquark component, the $[ud]$ diquark with isospin $I=0$ is
stable and keeps unchanged while the $[us]$ diquark is broken to
combine with the $\bar s$ to form either $K^+(u\bar
s)\Lambda([ud]s)$ or $\eta(s\bar s)p([ud]u)$.

The lighter $\Lambda^*(1405)1/2^-$ is also understandable in this
picture. Its dominant 5-quark configuration is $|[ud][us]\bar u>$
which is lighter than the corresponding 5-quark configuration
$|[ud][us]\bar s>$ in the $N^*(1535)1/2^-$.

From above results, we see that the diquark cluster picture for
the 5-quark components in baryons also gives a natural explanation
to the longstanding mass-reverse problem of $N^*(1535)$,
$N^*(1440)$ and $\Lambda^*(1405)$ resonances as well as the
unusual decay pattern of the $N^*(1535)$ resonance. However, if
this picture is correct, there should also exist the SU(3)
partners of the $N^*(1535)$ and $\Lambda^*(1405)$, {\sl i.e.}, an
additional $\Lambda^*~1/2^-$ around 1570 MeV, a triplet
$\Sigma^*~1/2^-$ around 1360 MeV and a doublet $\Xi^*~1/2^-$
around 1520 MeV \cite{zhusl}. There is no hint for these baryon
resonances in the PDG tables \cite{PDG}. Where are they? Here I
want to point out that there is indeed evidence for all of them in
the data of $J/\psi$ decays at BES. \vspace{-1cm}

\begin{figure}[htbp]
\hspace{1.5cm}\includegraphics[scale=0.3]{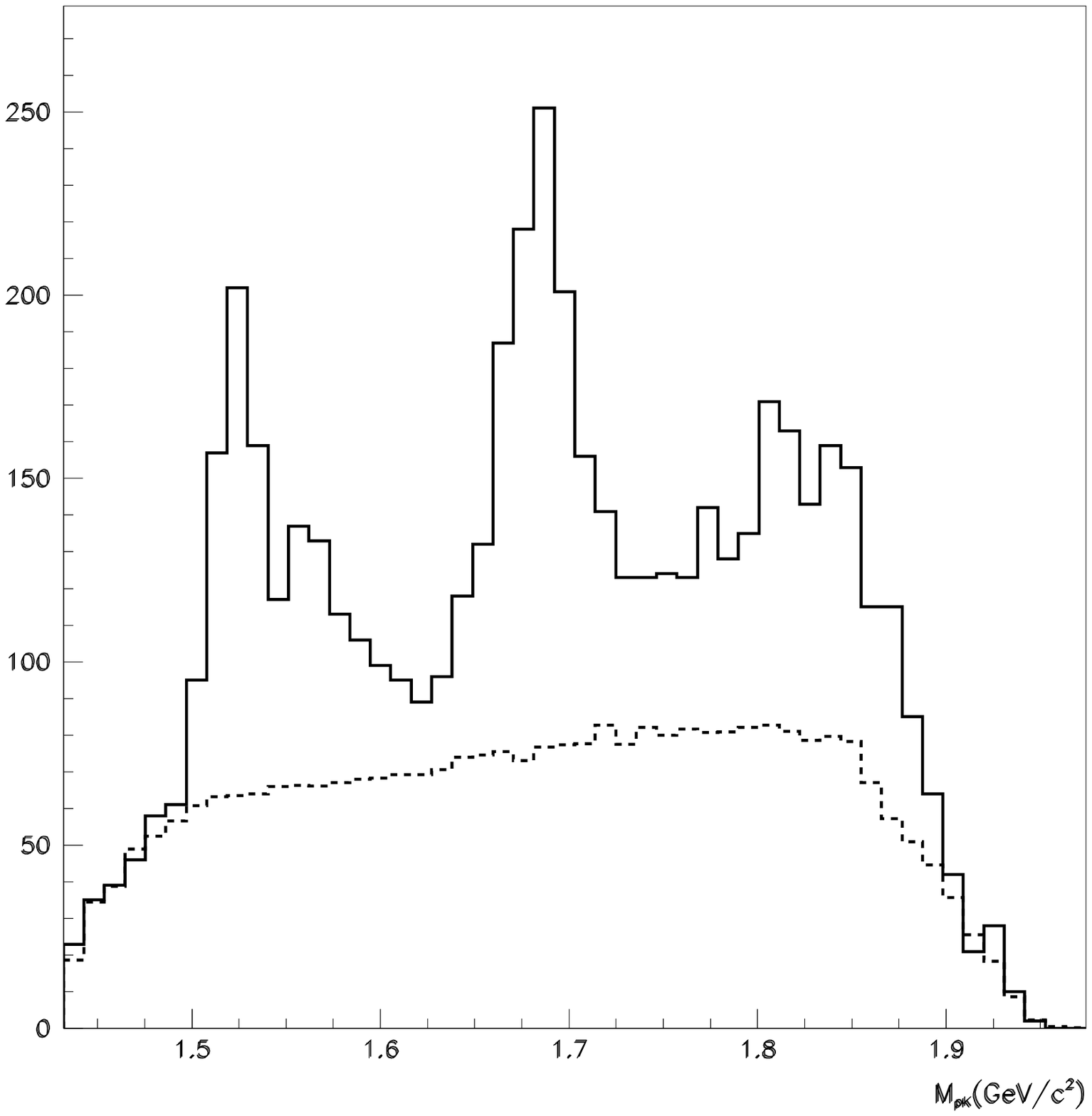}
\hspace{1cm}\includegraphics[scale=0.3]{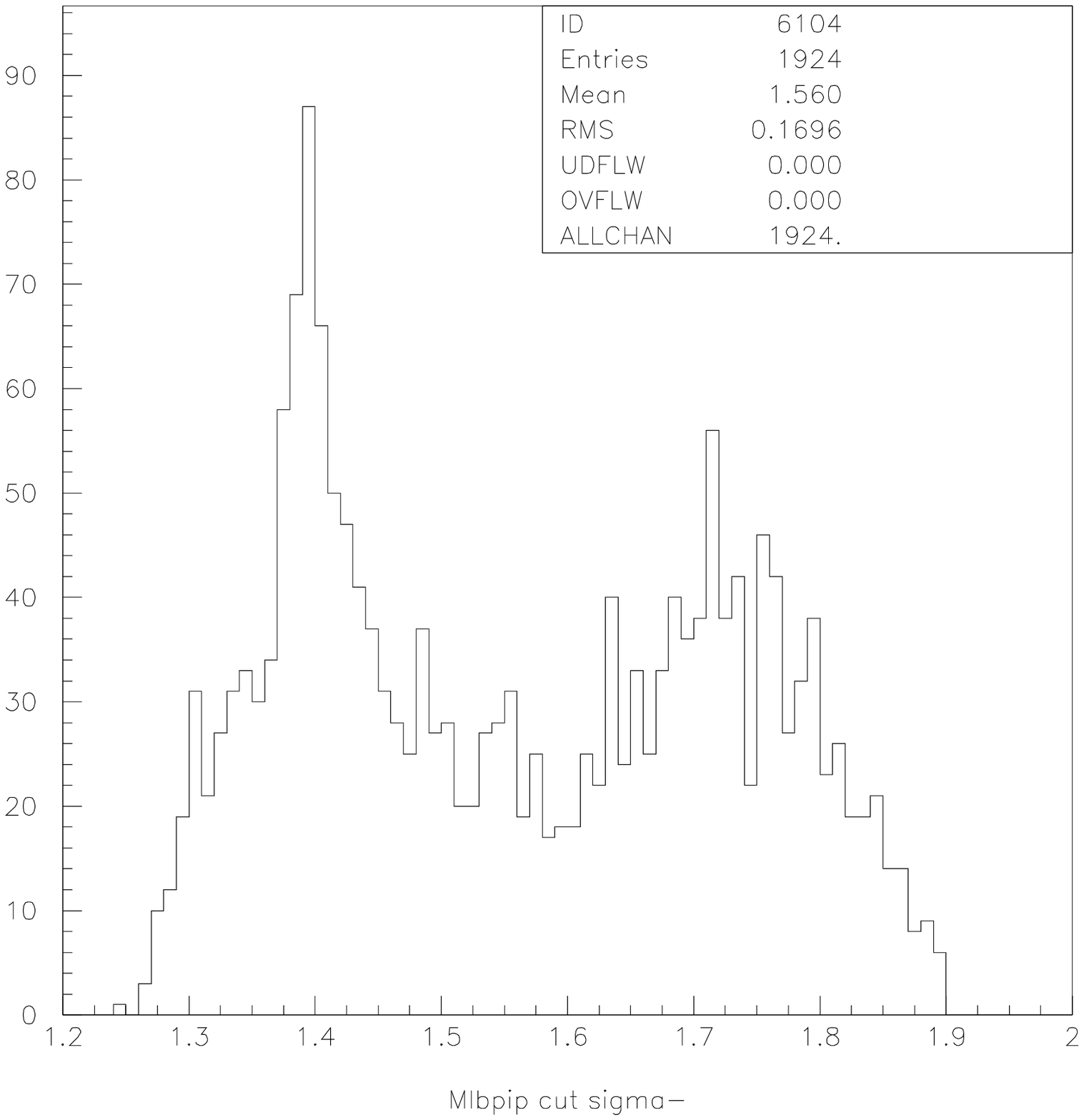} \caption{
\label{fig2} $pK$ invariant mass spectrum (left) for $J/\psi\to
pK^-\bar\Lambda$+c.c. and $\Lambda\pi$ invariant mass spectrum
(right) for $J/\psi\to\Lambda\bar\Sigma^+\pi^-$ from BES
\cite{zoubs4} }
\end{figure}
\vspace{-0.5cm}

Fig.~\ref{fig2} shows the $pK$ invariant mass spectrum (left) for
$J/\psi\to pK^-\bar\Lambda$+c.c. and $\Lambda\pi$ invariant mass
spectrum (right) for $J/\psi\to\Lambda\bar\Sigma^+\pi^-$ from BES
\cite{zoubs4}. In the $pK$ invariant mass spectrum, under the
narrow $\Lambda^*(1520)~3/2^-$ peak, there is a quite obvious
broader peak around 1570 MeV. Preliminary partial wave analysis
\cite{yanghx2001} gave its spin-parity as $1/2^-$. This
$\Lambda^*(1570)~1/2^-$ resonance fits in the new scheme for the
$1/2^-$ SU(3) baryon nonet very well. In the $\Lambda\pi$
invariant mass spectrum, under the $\Sigma^*(1385)~3/2^+$ peak,
there is also a broader peak around 1360 MeV. No partial wave
analysis has been performed for this channel yet. But there is a
good reason to reckon that there may be $1/2^-$ component
underneath the $\Sigma^*(1385)~3/2^+$ peak.

According to PDG \cite{PDG}, the branching ratios for
$J/\psi\to\bar\Sigma^-\Sigma^*(1385)^+$ and
$J/\psi\to\bar\Xi^+\Xi^*(1530)^-$ are $(3.1\pm 0.5)\times 10^{-4}$
and $(5.9\pm 1.5)\times 10^{-4}$, respectively. These two
processes are SU(3) breaking decays since $\Sigma$ and $\Xi$
belong to SU(3) $1/2^+$ octet while $\Sigma^*(1385)$ and
$\Xi^*(1530)$ belong to SU(3) $3/2^+$ decuplet. Comparing with the
similar SU(3) breaking decay $J/\psi\to\bar p\Delta^+$ with
branching ratio of less than $1\times 10^{-4}$ and the SU(3)
conserved decay $J/\psi\to\bar pN^*(1535)^+$ with branching ratio
of $(10\pm 3)\times 10^{-4}$, the branching ratios for
$J/\psi\to\bar\Sigma^-\Sigma^*(1385)^+$ and
$J/\psi\to\bar\Xi^+\Xi^*(1530)^-$ are puzzling too high. A
possible explanation for this puzzling phenomena is that there
were substantial components of $1/2^-$ under the $3/2^+$ peaks but
the two branching ratios were obtained by assuming pure $3/2^+$
contribution. This possibility should be easily checked with the
high statistics BESIII data in near future.

\section{Summary}

The empirical indications for a positive strangeness magnetic
moment and positive strangeness radius of the proton suggest that
the 5-quark components in baryons may be mainly in colored diquark
cluster configurations rather than in ``meson cloud''
configurations or in the form of a sea of quark-antiquark pairs.
The diquark cluster picture also gives a natural explanation for
the excess of $\bar d$ over $\bar u$ in the proton with a mixture
of $[ud][ud]\bar d$ component in the proton. More precise
measurements and analyses of the strange form factors are needed
to examine the relative importance of the meson-cloud components
and $q^2q^2\bar q$ components in the proton.

For excited baryons, the excitation energy for a spatial
excitation could be larger than to drag out a $q\bar q$ pair from
gluon field with the $q$ to form diquark cluster with a valence
quark. Hence the 5-quark components could be dominant for some
excited baryons.

The diquark cluster picture for the 5-quark components in baryons
also gives a natural explanation for the longstanding mass-reverse
problem of $N^*(1535)$, $N^*(1440)$ and $\Lambda^*(1405)$
resonances as well as the unusual decay pattern of the $N^*(1535)$
resonance with a large $|[ud][us]\bar u>$ component.

The diquark cluster picture predicts the existence of the SU(3)
partners of the $N^*(1535)$ and $\Lambda^*(1405)$, {\sl i.e.}, an
additional $\Lambda^*~1/2^-$ around 1570 MeV, a triplet
$\Sigma^*~1/2^-$ around 1360 MeV and a doublet $\Xi^*~1/2^-$
around 1520 MeV \cite{zhusl}. There is evidence for all of them in
the data of $J/\psi$ decays at BES, which should be examined by
high statistics data to be collected by BESIII in near future. One
may also search for these $1/2^-$ hyperon excited states under the
well known $\Sigma^*(1385)$, $\Lambda^*(1520)$ and $\Xi^*(1530)$
peaks in various other reactions, such as those at CEBAF and
Sping-8.

\section*{Acknowledgements}
I thank C.S.An, B.C.Liu and D.O.Riska for collaboration on
relevant issues.  The work is partly supported by CAS Knowledge
Innovation Project (KJCX2-SW-N02) and the National Natural Science
Foundation of China under grants Nos.10225525 \& 10435080.

\end{document}